\documentclass[prb,twocolumn,amsmath,amssymb,floatfix,superscriptaddress]{revtex4-1}

\usepackage{hyperref}
\usepackage{graphicx}
\usepackage{bm}
\usepackage{color}
\usepackage[applemac]{inputenc}
\usepackage{nicematrix}

\newcommand{\beq}{\begin{equation}}
\newcommand{\eeq}{\end{equation}}
\newcommand{\beqa}{\begin{eqnarray}}
\newcommand{\eeqa}{\end{eqnarray}}
\newcommand{\sdag}{{\dag}}

\begin{document}

\title{Deflation algorithm for the Green function of quasi-1D lattices}

\author{Pablo San-Jose}
\affiliation{Instituto de Ciencia de Materiales de Madrid, Consejo Superior de Investigaciones Cient\'{i}ficas (ICMM-CSIC), Madrid, Spain}
 
\date{\today}
 
\begin{abstract}
We derive a method to efficiently compute the Green function of on arbitrary Hamiltonians defined on semi-infinite and periodic quasi-one-dimensional lattices. Computing the Green function is the backbone of quantum transport, electronic structure or linear response computations. Our method constitutes a ``deflation optimization'' of a well established algorithm often used in quantum transport that is based on generalized Schur factorizations of the linearized quadratic eigenvalue equation for the transfer matrix. Our deflation optimization may greatly reduce the number of degrees of freedom that must be processed in the Schur factorization. Deflation must be supplemented by a Jordan-block reconstruction of generalized eigenvectors, also developed here in detail. The overhead of deflation plus reconstruction is minimal as compared to the typical reduction in factorization runtime. Furthermore, by avoiding inverses of ill-conditioned matrices, the algorithm remains numerically stable.
\end{abstract}

\maketitle

A basic pillar of computational condensed matter physics is the retarded Green function of electrons \cite{Economou:83,Abrikosov:75,Haug:08}. This quantity, denoted here as $G^r(t'-t)$ (in the time domain) or $G^r(\omega)$ (in the frequency domain), represents the causal propagation in a given time interval $t'-t$ of quantum particles or waves between two points in a stationary system. For numerical purposes, the system is often modeled using a tight-binding-like Hamiltonian on a lattice (constructed either ab-initio or phenomenologically). The non-equilibrium transport properties of the system without interactions or its spectral density, to name just two important examples, can be expressed fully in terms of $G^r$. The powerful Keldysh Green function formalism allows to cleanly extend non-interacting results to systems with interactions, in or out of equilibrium\cite{Keldysh:SPJ65,Kadanoff:62,Abrikosov:75,Rammer:RMP86,Meir:PRL92,Stefanucci:13}. Accurate and performant methods to compute $G^r$ are a critical part in all these techniques.

When computing the retarded Green function directly in the frequency domain $G^r(\omega)$ (the form most relevant to steady-state transport or spectral density simulations -- see e.g. Refs. \onlinecite{Gaury:PR14,Weston:PRB16} for alternative real-time approaches), numerical methods can be broadly classified into two groups: full-band and single-shot. In the first group one obtains $G^r(\omega)$ at all values of $\omega$ simultaneously. Prominent examples are the Tetrahedron Method \cite{Rath:PRB75,MacDonald:JPCSSP79,Molenaar:JPC82,Blochl:PRB94,Zaharioudakis:CPC05,Kawamura:PRB14}, that computes the non-interacting $G^r(\omega)$ from a sampling of the bandstructure of a periodic system, or the Kernel Polynomial Method\cite{Weisse:RMP06,Joao:RSOS20,Fan:PR21}, that uses Chebyshev expansion of functions of the system Hamiltonian, also without interactions. In both a price is paid in accuracy that can be reduced by increasing the sampling precision of the bandstructure or increasing the order of the expansion, respectively. In contrast, single-shot approaches \cite{Sanvito:PRB99,Rungger:PRB08} can typically compute $G^r(\omega)$ exactly (i.e. within machine precision), possibly in the presence of many-body self-energy corrections, at the expense of requiring one independent computation per frequency $\omega$. These are usually the preferred methods in normal transport and low-energy electronic structure simulations, wherein only electrons close to the Fermi energy are relevant. Of particular interest for mesoscopic transport are single-shot algorithms for periodic, quasi-one-dimensional (quasi-1D) systems, that allow to integrate out the semi-infinite quasi-1D leads connecting the system to electronic reservoirs \cite{Datta:97,Wimmer:09}.

\begin{figure}
   \centering
   \includegraphics[width=\columnwidth]{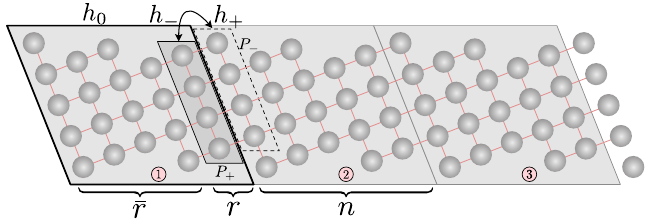} 
   \caption{Sketch of a generic periodic quasi-1D lattice. Each shaded circle represents an orbital, with red lines representing arbitrary hoppings. There are $n$ orbitals in each unit cell. $h_0$ is the $n\times n$ intracell Hamiltonian matrix. An $r$-dimensional orbital subspace $P_+$ in each unit cell (dark gray box) is coupled through $h_+$ to the neighboring unit cell on the right (intercell hoppings), while $h_-$, typically the adjoint of $h_+$, couples a subspace $P_-$ (dashed box) to the neighboring unit cell on the left. The self-energy in the first unit cell $\Sigma^r_{11}$ is concentrated on $P_+$. When computing the Green function between any two orbitals in the lattice, the deflation technique effectively eliminates the $\bar r$-dimensional subspace orthogonal to $P_+$ from the problem.}
   \label{fig:sketch}
\end{figure}

In this paper we present a ``deflation'' technique that can substantially speed up the performance of one of the most common single-shot method for periodic quasi-1D systems, dubbed here the Schur method\cite{Wimmer:09}, that is used in several numerical libraries for quantum transport\cite{Groth:NJOP14,Quantica:Z21}. The core of this method is a generalized Schur factorization (also known as QZ decomposition) of a linearized version of the Dyson equation for the transfer matrix. We show that in many cases of interest a potentially large subspace of electronic orbitals can be eliminated (``deflated'')  before the main Schur factorization step, which then becomes substantially faster due to its $O(n^3)$ complexity in the number $n$ of orbitals in the lead unit cell. The deflation pre-processing needs to be supplemented with a generalized eigenvector computation in a post-processing step that reconstructs any generalized eigenvectors that are lost to deflation. The overhead added by the pre-processing and post-processing steps are found to be typically negligible as compared to the runtime gains in the central Schur step.

\section{The Schur method}

Before developing the deflation and generalized eigenvector algorithms, we will first summarize the basic Schur method as developed in Ref. \onlinecite{Wimmer:09}. The Schur method solves, at each fixed $\omega$, the self-consistent Dyson equation $G^r(\omega)=g_0(\omega)+g_0(\omega)\Sigma^r(\omega)G^r(\omega)$ of a semi-infinite periodic lattice (a \emph{lead} in transport simulations, see Fig. \ref{fig:sketch}). Here $g_0(\omega) = (\omega+i\eta-h_0)^{-1}$ is the $n\times n$ Green function matrix of an isolated, $n$-orbital lead unit cell, of Hamiltonian $h_0$, and $\Sigma^r(\omega)$ is the self-energy due to the coupling $h_+$ ($h_-$) of each unit cell to its right (left) neighboring cells\footnote{For simplicity, and without loss of generality, we assume no intercell coupling beyond nearest-neighbors, as in practice these can always be eliminated by choosing a larger unit cell.}. If we project the Dyson equation onto the first (leftmost) unit cell of the semi-infinite lead, we have $G^r_{11}=g^r_0+g^r_0\Sigma^r_{11}G^r_{11}$ (we omit the $\omega$ dependences for brevity), or equivalently
\beq
\label{Dyson}
(g_0^{-1} - h_-G^r_{11}h_+)G^r_{11} = 1
\eeq
where $\Sigma^r_{11}=h_-G^r_{11}h_+$ is the self-energy induced on the first unit cell by the remaining unit cells, and $G^r_{11}$ denotes the retarded Green function matrix between any two sites in the first unit cell. We rewrite this equation by multiplying with $h_+$ on the right and by defining the retarded transfer matrix $T^r = G^r_{11}h_+$,
\beq
\label{DysonT}
h_+ - g_0^{-1}T^r + h_-(T^r)^2 = 0
\eeq
If we assume $T^r$ is diagonalizable (which is not the case in general, as will be discussed below), we can write
\beq
\label{T}
T^r=G^r_{11}h_+=\Phi^r\Lambda^r\left(\Phi^r\right)^{-1},
\eeq
where $\Phi^r$ is an $n\times n$ invertible matrix with the proper eigenvectors $\phi_i$ of $T^r$ as columns, and $\Lambda^r$ is a diagonal matrix with eigenvalues $\lambda_i$ in the diagonal, so that $T^r\phi_i = \lambda_i\phi_i$ for each eigenpair $\phi_i$, $\lambda_i$. By combining Eqs. \eqref{DysonT} and \eqref{T}, we see that all eigenpairs $\phi, \lambda$ satisfy a quadratic eigenvalue equation
\beq
\label{lin}
(h_+ - g_0^{-1}\lambda + h_-\lambda^2)\phi = 0
\eeq

The transfer matrix also needs to satisfy \emph{retarded} boundary conditions, which corresponds to a finite $\lim_{m\to\infty} (T^r)^m$. This restricts the allowed (complex) eigenvalues $\lambda$ to lie on the unit circle or inside, $|\lambda|\leq 1$. 
If we rewrite $\lambda = e^{ika_0}$, where $a_0$ is the lead lattice period and $k$ is a real or complex wavevector, we may interpret Eq. \eqref{lin} above as the Bloch equation for propagating ($|\lambda|=1$), right-evanescent ($|\lambda|<1$) or left-evanescent ($|\lambda|>1$) solutions at energy $\omega$ on an infinite version (instead of semi-infinite) of the lead lattice. To preserve causality we must restrict $\Phi^r$ to solutions with $|\lambda|\leq 1$, and among propagating ones ($|\lambda|=1$) keep only those with \emph{positive} group velocity. These can be shown \cite{Wimmer:09} to become evanescent when $\omega$ acquires a small and positive imaginary part $i\eta$, as in the definition of $g_0(\omega)$. In practice, then, we define $\Phi^r$ simply as the set of eigenvectors with $|\lambda|<1$ when $\eta \to 0^+$. \footnote{More advanced versions of the algorithm often use the velocity operator to disentangle retarded and advanced propagating solutions on the unit circle without resorting to complex $\omega$, but we will not discuss these details here.} Assuming we can find $n$ such linearly independent retarded solutions (this is possible when $T^r$ is diagonalizable), $\Phi^r$ will be a full-rank, invertible matrix, which allows to build $T^r$ using Eq. \eqref{T} and solves the problem of finding $G^r_{11}$ through Eq. \eqref{Dyson},
\beqa
\label{G}
G^r_{11} &=& (g_0^{-1} - \Sigma^r_{11})^{-1} \\
\Sigma^r_{11} &=& h_-T^r = h_- \Phi^r\Lambda^r\left(\Phi^r\right)^{-1}\nonumber
\eeqa
With $G^r_{11}$ we can also reconstruct $G^r_{NM}$ between any two unit cells $N$ and $M$, as explained in Appendix \ref{ap:GNM}.

The problem then reduces to finding all retarded solutions of Eq. \eqref{lin}. A powerful and quite common approach is to linearize the $n\times n$ quadratic eigenvalue problem into a $2n\times 2n$ generalized but linear eigenvalue problem $A\psi = \lambda B\psi$. This is done using one of several possible linearization transformations. The so-called first-companion linearization is probably simplest, and takes the following form
\beq
\label{pencil}
\Bigg[\overbrace{\left(\begin{array}{cc}
0 & 1\\ -h_+ & g_0^{-1}
\end{array}\right)}^{A}
-
\lambda\overbrace{\left(\begin{array}{cc}
1 &0 \\ 0 &h_-
\end{array}\right)}^{B}\Bigg]
\overbrace{\left(\begin{array}{c}
\phi \\ \chi
\end{array}\right)}^{\psi} = (A-\lambda B)\psi = 0
\eeq
Each advanced $|\lambda|>1$ or retarded $|\lambda|<1$ solution to the above corresponds to one and only one solution of the original quadratic eigenvalue problem, where $\chi = \lambda\phi$. At a low level, numerical libraries such as LAPACK \cite{Anderson:99} compute the eigenpairs $\lambda, \psi$ in two steps. First, a generalized Schur (or QZ) factorization of $A,B$,
\beq
\label{Schur}
A-\lambda B = Q(S_A-\lambda S_B)Z^\sdag,
\eeq
does the heavy lifting of finding unitary $2n\times 2n$ matrices $Q$ and $Z$ that transform  $A$ and $B$ into \emph{upper triangular} matrices $S_A$ and $S_B$, respectively. The diagonals $\alpha_i$ and $\beta_i$ of $S_A$ and $S_B$ can be shown to yield the eigenvalues $\lambda_i = \alpha_i/\beta_i$. The Schur factorization is numerically stable, and once obtained, efficient routines exist that can transform it into an equivalent Schur form with reordered diagonals of $S_A, S_B$, wherein retarded eigenvalues ($|\alpha_i| < |\beta_i|$) appear before advanced eigenvalues. Then, a backsubstitution stage is used for each $\lambda$ to find the $\psi$ eigenvectors one by one. The resulting retarded and advanced eigenvectors $\psi$ can be written as the columns of the matrix $\Psi = ZR$, where $R$ is the backsubstitution upper-triangular matrix. If the factorization was reordered with retarded solutions coming first, we have
\beqa
\label{Psi}
\Psi
=
\left(\begin{array}{cc}
\Phi^r & \Phi^a \\ \Phi^r\Lambda^r & \Phi^a\Lambda^a
\end{array}\right)
=
\left(\begin{array}{cc}
Z_{11} & Z_{12}\\ Z_{21} & Z_{22}
\end{array}\right)
\left(\begin{array}{cc}
R_{11} & R_{12}\\ 0 & R_{22}
\end{array}\right),
\eeqa
where we have also included the advanced $\Phi^a$, $\Phi^a\Lambda^a$ blocks with $|\lambda|>1$. Note that $\Phi^r = Z_{11}R_{11}$ and $\Phi^r\Lambda^r=Z_{21}R_{11}$. This expression allows us to interpret $Z_{11}$ as the basis of the invariant subspaces spanned by subsets of eigenstates $\Phi^r$, and $R_{11}$ as the eigenstate coordinates in this basis. (Similarly for $Z_{21}$ and states $\Phi^r\Lambda^r$.)
Crucially, the $T^r$ matrix is independent of the coordinate matrix $R$, since
\beq
\label{T2}
T^r = \Phi^r\Lambda^r\left(\Phi^r\right)^{-1} = Z_{21}R_{11}R_{11}^{-1}Z_{11}^{-1} = Z_{21}Z_{11}^{-1}
\eeq
This is a very significant advantage of the Schur method, and a common theme throughout this work. We can construct $T^r$ matrix, and hence the Green function, without needing to actually compute matrix $R$ or eigenstates $\Phi^r$. Only the bases $Z_{11}$ and $Z_{21}$ of the invariant subspaces is needed. This is an important advantage numerically, since it also bypasses the need to invert $\Phi^r$, which can be numerically unstable (note that since $T^r$ is non-Hermitian, so its eigenstates can be almost parallel and even coalesce, yielding ill-conditioned or even singular $R_{11}$). In contrast to $R_{11}^{-1}$, the inverse $Z_{11}^{-1}$ is typically found to be well conditioned, since it is an $n\times n$ block of the $Z$ unitary matrix. A further, often overlooked advantage is that Eq. \ref{T2} gives the correct retarded solution of Eq. \eqref{DysonT} even if the resulting $T^r$ is \emph{defective} (non-diagonalizable), i.e. when $T^r$ has only $r<n$ linearly independent proper eigenvectors, making $R$ strictly rank-deficient and thus non-invertible.

\section{The deflation technique}
\label{sec:deflation}

The main practical drawback of the Schur method as presented here is the fact that its theoretical computational complexity grows with the cube of the number of orbitals $n$, which can prove to be a challenge for large unit cells. The deflation technique presented in this section optimizes the Schur factorization step by removing all $\lambda=\infty$ advanced eigenvectors and all $\lambda=0$ retarded eigenvectors from the problem. Note that neither of these contribute to $T^r$ as given in Eq. \eqref{T}. The existence of retarded $\lambda=0$ ($\lambda=\infty$) eigenvectors is a consequence matrix $h_+$ ($h_-$) having a non-empty kernel in general. Indeed, if a non-zero vector $\phi_0$ belongs to $\mathrm{ker}(h_+)$ it will be a $\lambda=0$ eigenvector, since by definition $T^r\phi_0 = G^r_{11}h_+\phi_0=0$. Geometrically, any orbitals in an $n$-orbital unit cell not directly coupled by $h_+$ to the neighboring unit cell on the right will belong to $\mathrm{ker}(h_+)$, see Fig. \ref{fig:sketch}. By removing all $\bar{r} = n-r = \mathrm{dim}\,\mathrm{ker}(h_+)$ solutions with $\lambda=0$ beforehand, we may transform the $2n\times 2n$ Schur factorization step of Eq. \eqref{Schur} into a $2r\times 2r$ factorization, roughly yielding a $\sim(r/n)^3$ reduction in runtime. The deflation technique is a prescription for building a smaller $2r\times 2r$ linear pencil $A_2 -\lambda B_2$ which however shares the same spectrum as pencil $A-\lambda B$ in Eq. \eqref{pencil} except for any $\lambda=0$ and $\lambda=\infty$ solutions. The only requirement for the deflation technique in this section to work is that $\mathrm{ker}(h_+) = \mathrm{ker}(h_-^\sdag)$, which is guaranteed for Hermitian lattice Hamiltonians with $h_+ = h_-^\sdag$. In Appendix \ref{ap:quadeig} we sketch a more elaborate deflation procedure (a variation of the so-called \verb|quadeig| algorithm \cite{Hammarling:ATMS13,Drmac:19}) that works also with arbitrary $h_\pm$. Throughout this section we also assume that $T^r$ is non-defective. The fully general algorithm for the defective case will be completed in Sec. \ref{sec:defective}.

The first step is to find orthonormal bases $\bar{P}_+$ and $P_+$ of $\mathrm{ker}(h_+)$ and its orthogonal complement, respectively. The $\bar{P}_+$ matrix is $\bar r\times n$ , and $P_+$ is $r\times n$, with $r+\bar r = n$, so that together they form a unitary $n\times n$ matrix $Q_+ = (P_+ \bar P_+)$. It can be efficiently computed using a pivoted LQ decomposition of $h_+$, \footnote{Pivoted LQ decomposition algorithms are implemented in most numerical linear algebra libraries. The pivoted LQ of a matrix $m$ can also be computed as the adjoint of the more common pivoted QR decomposition, applied to the adjoint of $m$. The determination of the zeros in the $L$ matrix requires some numerical tolerance criterion, which is implicit throughout this work. In practical implementations a good choice is to consider an entry of $L$ zero if its absolute value is below the square root of the floating point precision around one.}
\beq
\label{h+}
h_+ = 
\begin{pNiceMatrix}[
	first-row,
	code-for-first-row = \scriptstyle,
	code-for-last-col = \scriptstyle]
r & \bar r\\
L_+ & 0
\end{pNiceMatrix}
\begin{pNiceMatrix}[
	first-row,
	code-for-first-row = \scriptstyle,
	code-for-last-col = \scriptstyle]
n \\ P_+^\sdag\\ \bar P_+^\sdag
\end{pNiceMatrix} = 
\begin{pNiceMatrix}
L & 0
\end{pNiceMatrix}
Q_+^\sdag,
\eeq
so that
\beqa
Q_+^\sdag
h_+Q_+ &=& 
\begin{pNiceMatrix}
V_+ & 0\\
\bar V_+ & 0
\end{pNiceMatrix}
\\
\label{h-}
Q_+^\sdag
h_-Q_+ &=& 
\begin{pNiceMatrix}
V_- & \bar V_-\\
0 & 0
\end{pNiceMatrix}
\\
Q_+^\sdag
g_0^{-1}Q_+ &=& 
\begin{pNiceMatrix}
g_{0rr}^{-1} & g_{0r\bar r}^{-1}\\
g_{0\bar rr}^{-1} & g_{0\bar r\bar r}^{-1}
\end{pNiceMatrix}
\eeqa
The first companion linearization in Eq. \eqref{pencil} is transformed under the unitary transformation \[Q_1 = \begin{pNiceMatrix}
Q_+ & 0 \\ 0 & Q_+
\end{pNiceMatrix}\] into $A_1-\lambda B_1 = Q_1^\sdag(A-\lambda B)Q_1$, and the expanded eigenvectors into $\psi_1 = Q_1^\sdag \psi = (\phi_r,\phi_{\bar r},\chi_r, \chi_{\bar r})^{T}$, where $\phi_r = P^\sdag_+\phi$ represents the projection of the eigenstate on the right `surface' of the unit cell, and $\phi_{\bar r} = \bar P^\sdag_+\phi$ its orthogonal. The eigenvalue equation $(A_1-\lambda B_1)\psi_1=0$ then reads
\beq
\Bigg[
\overbrace{
\begin{pNiceMatrix}
0 & 0 & 1 & 0\\ 0 & 0 & 0 & 1\\
-V_+ & 0 & g_{0rr}^{-1} & g_{0r\bar r}^{-1}\\
-\bar V_+ & 0 & g_{0\bar rr}^{-1} & g_{0\bar r\bar r}^{-1}
\end{pNiceMatrix}}^{A_1}
-\lambda
\overbrace{\begin{pNiceMatrix}
1 & 0 & 0 & 0\\ 0 & 1 & 0 & 0\\
0 & 0 & V_- & \bar V_-\\
0 & 0 & 0 & 0
\end{pNiceMatrix}}^{B_1}
\Bigg]
\overbrace{\begin{pNiceMatrix}
\phi_r \\ \phi_{\bar r} \\ \chi_r \\ \chi_{\bar r}
\end{pNiceMatrix}}^{\psi_1} = 0
\eeq
In this basis we immediately see that only the $\phi_r$, $\chi_r$ and $\chi_{\bar r}$ eigenvector components are coupled, with $\phi_{\bar r}$ decoupling away, and given simply by $\phi_{\bar r} = \lambda^{-1}\chi_{\bar r}$ once $\chi_{\bar r}$ is known (unless $\lambda = 0$). We are thus left with 
\beq
\Bigg[
\overbrace{\begin{pNiceMatrix}
0 & 1 & 0\\
-V_+ & g_{0rr}^{-1} & g_{0r\bar r}^{-1}\\
-\bar V_+ & g_{0\bar rr}^{-1} & g_{0\bar r\bar r}^{-1}
\end{pNiceMatrix}}^{\tilde A_1}
-\lambda
\overbrace{\begin{pNiceMatrix}
1 & 0 & 0\\ 
0 & V_- & \bar V_-\\
0 & 0 & 0
\end{pNiceMatrix}}^{\tilde B_1}
\Bigg]
\overbrace{\begin{pNiceMatrix}
\phi_r\\ \chi_r \\ \chi_{\bar r}
\end{pNiceMatrix}}^{\tilde \psi_1} = 0
\eeq
By casting the eigenproblem in this form we have implicitly dropped all $\lambda = 0$ solutions, which have a finite $\phi_{\bar r}$ with all other components equal to zero.
From the last block-row we now see that any $\tilde \psi_1 =(\phi_r,\chi_r, \chi_{\bar r})^{T}$ eigenvector with $\lambda\neq\infty$ must belong to the kernel of $C_2 = \begin{pNiceMatrix}-\bar V_+ & g_{0\bar rr}^{-1} &g_{0\bar r\bar r}^{-1}\end{pNiceMatrix}$. $C_2$ is a matrix of size $\bar r\times (n+r) = (n-r)\times (n+r)$, so it kernel is at least of dimension $2r$. We compute an orthonormal basis $\bar P_2$ of $\mathrm{ker}(C_2)$ through another LQ factorization
\beq
\label{P2}
C_2 =\begin{pNiceMatrix}-\bar V_+ & g_{0\bar rr}^{-1} &g_{0\bar r\bar r}^{-1}\end{pNiceMatrix} =
\begin{pNiceMatrix}[
	first-row,
	code-for-first-row = \scriptstyle]
\bar r & 2r\\
L_2 & 0
\end{pNiceMatrix}
\begin{pNiceMatrix}[
	first-row,
	code-for-first-row = \scriptstyle]
n+r \\ P_2^\sdag \\ \bar P^\sdag_2
\end{pNiceMatrix}
\eeq
We define $\psi_2 = \bar P^\sdag_2\tilde \psi_1$ as the $2r$ components of $\tilde\psi_1$ within $\mathrm{ker}(C_2)$. The projection $(\tilde A_1-\lambda \tilde B_1)\bar P_2 = A_2 - \lambda B_2$ gives our final, fully deflated $2r \times 2r$ pencil, with
\beqa
A_2 &=& \begin{pNiceMatrix}
0 & 1 & 0\\
-V_+ & g_{0rr}^{-1} & g_{0r\bar r}^{-1}
\end{pNiceMatrix} \bar P_2 \nonumber\\
B_2 &=& \begin{pNiceMatrix}
1 & 0 & 0\\ 
0 & V_- & \bar V_-
\end{pNiceMatrix} \bar P_2
\label{deflated}
\eeqa
For each of its proper eigenvectors $\psi_2$ satisfying
\beq
\label{defpencil}
(A_2 - \lambda B_2)\psi_2 = 0
\eeq
we can reconstruct a proper eigenvector $\phi$ of the original quadratic eigenvalue equation by reverting the two transformations $Q_1$ and $\bar P_2$. However, it is better to proceed like in the original Schur algorithm. We perform a generalized Schur decomposition, although this time on the smaller ($2r\times 2r$) deflated $A_2$, $B_2$,
\beq
\label{defSchur}
A_2-\lambda B_2 = Q(S_{A_2}-\lambda S_{B_2})Z^\sdag
\eeq
and proceed exactly like in Eq. \eqref{Psi}. Using the relation $\tilde\Psi_1 =\bar P_2 \Psi_2$ between the $\tilde\psi_1$ and $\psi_2$ eigenvector matrices, we arrive at
\beq
\label{Psi2}
\tilde\Psi_1
=
\begin{pNiceMatrix}
\Phi_r^r & \Phi_r^a \\
\Phi_r^r\Lambda^r & \Phi_r^r\Lambda^a \\
\Phi_{\bar r}^r\Lambda^r & \Phi_{\bar r}^r\Lambda^a
\end{pNiceMatrix}
=
\begin{pNiceMatrix}
P^\sdag_+\Phi^r & P^\sdag_+\Phi^a \\
Q^\sdag_+\Phi^r\Lambda^r & Q^\sdag_+\Phi^a\Lambda^a
\end{pNiceMatrix}
=
\bar P_2 Z R
\eeq
Here we have used the fact that the columns of $\Phi_r^r$ are the projection of the retarded eigenvectors on the orthogonal complement of $\mathrm{ker}(h_+)$, i.e. $\phi_r=P_+^\sdag\phi$. Once more $R$ is a certain upper-triangular matrix resulting from backsubstitution, this time of size $2r\times 2r$, and 
\beq
\label{Ztilde}
\tilde Z = \bar P_2 Z = 
\begin{pNiceMatrix}[
	first-row, first-col,
	code-for-first-row = \scriptstyle,
	code-for-first-col = \scriptstyle]
& r & r\\
r & \tilde Z_{11} & \tilde Z_{12}\\
n & \tilde Z_{21} & \tilde Z_{22}\\
\end{pNiceMatrix}
\eeq
is a matrix of size $(r+n)\times 2r$. We finally get
\beqa
\label{Z11}
P^\sdag_+\Phi^r &=& \tilde Z_{11}\tilde R_{11} \\ 
\Phi^r\Lambda^r &=& Q_+\tilde Z_{21}\tilde R_{11}\nonumber
\eeqa
Note that we cannot extract $\Phi^r$ from the above equation, since unlike the full $Q_+$, $P^\sdag_+$ is not a unitary matrix. However, we now show that this is enough to reconstruct $T^r$. If we separate, in Eq. \eqref{T2}, the $\lambda = 0$ retarded eigenvector matrix (denoted by $\Phi^r_0$, whose columns live in $\mathrm{ker}(h_+)$, so that $P^\sdag_+\Phi^r_0=0$) from the $\lambda\neq 0$ eigenvector matrix (denoted by $\Phi^r$), $T^r$ takes the form
\beqa
\label{T3}
T^r &=& \begin{pNiceMatrix} \Phi^r \Lambda^r& 0\end{pNiceMatrix}
\begin{pNiceMatrix} \Phi^r & \Phi^r_0\end{pNiceMatrix}^{-1}
\nonumber\\
&=& 
\begin{pNiceMatrix} \Phi^r \Lambda^r& 0\end{pNiceMatrix}
\Bigg[
\overbrace{\begin{pNiceMatrix} P_+ & \bar P_+\end{pNiceMatrix}}^{Q_+}
\overbrace{\begin{pNiceMatrix} P^\sdag_+\Phi^r& 0 \\ \bar P^\sdag_+\Phi^r & \bar P^\sdag_+ \Phi^r_0\end{pNiceMatrix}}^{Q^\sdag_+\begin{pNiceMatrix} \Phi^r & \Phi^r_0\end{pNiceMatrix}}\Bigg]^{-1} 
\nonumber \\
&=&\Phi^r \Lambda^r \left(P^\sdag_+\Phi^r\right)^{-1}P^\sdag_+
\eeqa
where the $r\times r$ matrix $P^\sdag_+\Phi^r = \tilde Z_{11}\tilde R_{11}$ is invertible by assumption (non-defective $T^r$), although $\tilde R_{11}$ may be very ill-conditioned. Like in the non-deflated Schur algorithm, this ill-conditioned $\tilde R_{11}$ factor drops out from $T^r$, which  finally reads
\beq
\label{Tfinal}
T^r = Q_+\tilde Z_{21} \tilde Z_{11}^{-1} P^\sdag_+
\eeq
This relation is the main result of this section. To summarize, the deflation algorithm has six steps: (1) LQ-decompose $h_+$ to obtain basis $Q_+$ and its $P^\sdag_+$ subspace, (2) build the $C_2$ matrix for a given $\omega$, (3) LQ-decompose $C_2$ as in Eq. \eqref{P2} to obtain $\bar P_2$, (4) build the deflated $A_2$, $B_2$ matrices using Eq. \eqref{deflated} and $\bar P_2$, (5) perform the generalized Schur decomposition of Eq. \eqref{defSchur}, and reorder it so that the retarded $|\lambda|<1$ eigenvalues come first, and (6) extract the $\tilde Z_{11}$ and $\tilde Z_{21}$ blocks in Eq. \eqref{Ztilde} using $\bar P_2$ and the Schur decomposition to insert in Eq. \eqref{Tfinal}.

\section{Defective case and generalized eigenvectors}
\label{sec:defective}

Equation \eqref{Tfinal} solves the problem of computing $T^r$, $\Sigma^r_{11}$ and $G^r_{11}$ in those cases where $\tilde Z_{11}$ is a full-rank $r\times r$ matrix. This happens in the simplest cases, where a full basis of proper retarded eigenstates of $T^r$ exist, $r$ of them with $0<|\lambda|<=1$ and $\bar r = n - r$ with $\lambda=0$. In this case all $\alpha_i$ entries in the diagonal of $S_{A_2}$ are non-zero, and the discussion to follow becomes unnecessary.

More generally, however, $T^r$ may be non-diagonalizable, which means that a basis of proper eigenvectors of $T^r$ does not exist. This implies that $\tilde Z_{11}$ turns out rank-deficient within numeric precision, or at least very ill-conditioned, and it is thus not possible to invert it in Eq. \eqref{Tfinal}. This is in contrast to the non-deflated $Z_{11}$, which is found to be reliably well-conditioned. In this section we show how to modify Eq. \eqref{Tfinal} to deal with this situation.

Any matrix, even if it is non-diagonalizable, can be expressed in terms of a full basis composed of a set of proper ($\Phi^r$) and a set of \emph{generalized} ($\Phi^g$) eigenvectors as follows
\beq
\label{Jordan}
T^r =
\begin{pNiceMatrix} \Phi^r & \Phi^g \end{pNiceMatrix}
\begin{pNiceMatrix} \Lambda^r & 0 \\ 0 & J^r \end{pNiceMatrix}
\begin{pNiceMatrix} \Phi^r & \Phi^g \end{pNiceMatrix}^{-1}
\eeq
where $J^r$ is a quasi-diagonal matrix composed of Jordan blocks along its diagonal. Each Jordan block $J_\lambda^r$ in $J^r$ is associated to an eigenvalue $\lambda$ that fills the the diagonal of the block, with ones above the diagonal and zeros elsewhere. Hence a block $J_\lambda^r$ of size $m$ will be nilpotent respect to $\lambda$ with degree $m-1$,
\beq
\label{nilpotent}
(J_\lambda^r - \lambda\,1_{m\times m})^{m-1} = 0.
\eeq

The deflation procedure explained in the preceding section eliminates both proper and generalized $\lambda=0$ eigenvectors. Unlike the former, generalized $\lambda=0$ eigenvectors do enter the expression for $T^r$ because $J_0^r$ is non-zero. It is therefore necessary, in general, to supplement the deflation procedure with an explicit reconstruction of the eliminated generalized eigenvectors $\Phi^g_0$ corresponding to $\lambda=0$. This allows us to write the generally valid expression
\beq
\label{Sigma}
\Sigma^r_{11} =h_-T^r = \begin{pNiceMatrix}h_-\Phi^r \Lambda^r & h_-\Phi^g_0 J^r_0\end{pNiceMatrix}
\left[P^\sdag_+\begin{pNiceMatrix}\Phi^r & \Phi^g_0\end{pNiceMatrix}\right]^{-1}P^\sdag_+
\eeq
which is obtained like in Eq. \eqref{T3} but using the generalized expansion in Eq. \eqref{Jordan}.
Here $\Phi^r$ is the matrix of $r-r_0$ proper retarded eigenvectors with \emph{non-zero} $\lambda$, and $\Phi^g_0$ are $r_0$ generalized eigenvectors corresponding to $\lambda=0$, which together complete a basis for the $r$-dimensional subspace spanned by $P_+$. 

We now describe an efficient method to compute $\Sigma^r_{11}$ in Eq. \eqref{Sigma} above by reconstructing the required information of the $\lambda=0$ generalized eigenvectors $\Phi^g_0$ that were eliminated by deflation. Note that we have intentionally chosen to target $\Sigma^r_{11}$ instead of $T^r$ this time. The reason is that it is more efficient to compute $h_-\Phi^g_0 J^r_0$ than $\Phi^g_0 J^r_0$ using the method presented below, and $\Sigma^r_{11}$ is actually enough to compute $G^r_{11}$ using Eq. \eqref{G}. In this sense $\Sigma^r_{11}$ is a more fundamental quantity than the transfer matrix.

The starting point for the method is the nilpotent property in Eq. \eqref{nilpotent}. Since 
\beq
\label{Phig}
T^r\Phi^g_0 = \Phi^g_0 J_0^r,
\eeq
Eq. \eqref{nilpotent} implies that
\beq
(T^r)^m\Phi^g_0 = 0
\eeq
for some $m>1$. According to Eq. \eqref{GN1} this is equivalent to $G^r_{m1}h_+\Phi^g_0 = 0$, where $G^r_{NM}$ is a propagator between unit cells $M$ and $N$. 

Let us take a moment to reflect on this result. Each column of $\Phi^g_0$ corresponds to a state $\phi^g_0$ such that if we propagate $h_+\phi^g_0$ from the first unit cell of a semi-infinite lattice to cell $m>1$, it vanishes. It thus corresponds to a peculiar surface state that decays into the bulk of the semi-infinite lattice, but unlike proper evanescent eigenvectors that decay exponentially, it vanishes \emph{exactly} a finite number of unit cells away from the surface. $\phi^g_0$ should moreover not be a proper $\lambda=0$ eigenstate, so it should not vanish under $T^r$, i.e. $T^r\phi^g_0\neq 0$. Our task is then to find the matrix $\Phi^g_0$ of all such vectors, together with their corresponding $\Phi^g_0 J_0^r = T^r\Phi^g_0$. 

To do this efficiently we rely on a key observation: since propagating a vector $h_+\phi^g_0$ in the first unit cell by a number $m>1$ of unit cells into the semi-infinite bulk makes it vanish exactly, it will also vanish exactly if we replace the semi-infinite lattice with a finite lattice containing only $m$ unit cells. Namely, instead of using $G^r_{m1}h_+\Phi^g_0 = 0$ we can also compute $\Phi^g_0$ and $\Phi^g_0J^r_0$ by using $g^m_{m1}h_+\Phi^g_0 = 0$ and $\Phi^g_0J^r_0 = g^m_{11}h_+\Phi^g_0$, where $g^m_{m1}$ and $g^m_{11}$ denote the propagators between the ends of a finite lattice with $m$ unit cells. This observation is useful because we do not know the $G^r_{m1}$ of the semi-infinite lattice (we are actually computing $\Phi^g_0$ to be able to build it), but computing $g^m_{m1}$ is actually easy using the equations of the recursive Green function method \cite{Thouless:JPCSSP81}, which can be used to relate propagators in a finite lattice of increasing length as follows,
\beqa
\label{recursive}
g^m_{11} &=& (g_0^{-1}-h_-g_{11}^{m-1}h_+)^{-1}\\
g^m_{m1} &=& g^{m-1}_{m-1,1}h_+(g_0^{-1}-h_-g_{11}^{m-1}h_+)^{-1}\nonumber
\eeqa
These recursive relations allow us to find the required $g^m_{11}$ and $g^m_{m1}$ for increasing $m$, starting from $g^1_{11} = g_0$. 

The $g^m_{11}$ and $g^m_{m1}$ are (potentially large) $n\times n$ matrices. To keep the overall algorithm fast, we also `deflate' the recursion by integrating out all orbitals outside the $P_+$ subspace. We can do this by multiplying Eqs. \eqref{recursive} with $P_+^\sdag h_-$ on the left and $h_+P_+$ on the right, so instead of $g^m_{11}$ and $g^m_{m1}$ we recursively compute the $r\times r$ matrices
\beqa
\tilde\Sigma_{11}^m &=& P_+^\sdag h_-g^m_{11}h_+P_+ \\
\tilde\Sigma_{m1}^m &=& P_+^\sdag h_-g^m_{m1}h_+P_+.
\eeqa
By using the additional $r\times r$ matrices
\beqa
\label{tilde}
\tilde g_0 &=& P_+^\sdag g_0 P_+\\
\tilde t_+ &=& P_+^\sdag g_0 h_+ P_+\nonumber\\
\tilde t_- &=& P_+^\sdag h_- g_0 P_+\nonumber\\
\tilde \Sigma_0 &=& P_+^\sdag h_+g_0h_-P_+\nonumber
\eeqa
we can deflate the original recursion relations into 
\beqa
\label{recursive2}
\tilde\Sigma_{11}^m &=& \tilde \Sigma_0 + \tilde t_-\tilde\Sigma^{m-1}_{11}\left(1-\tilde g_0\tilde\Sigma^{m-1}_{11}\right)^{-1}\tilde t_+\\
\tilde\Sigma_{m1}^m &=& \tilde\Sigma_{(m-1),1}^{m-1}\left(1-\tilde g_0\tilde\Sigma^{m-1}_{11}\right)^{-1}\tilde t_+
\eeqa
with seed $\tilde \Sigma_{11}^1 = \tilde\Sigma_0$.

For Eq. \eqref{Sigma} we want all projected solutions $P_+^\sdag\Phi_0^g$ such that $g^m_{m1} h_+ \Phi_0^g=0$. These are all states in $\mathrm{ker}(\tilde\Sigma^m_{m1})$. We compute a basis of this kernel using a pivoted LQ factorization as in previous occasions,
\beq
\label{sigmakernel}
\tilde\Sigma^m_{m1} = 
\begin{pNiceMatrix}
L^g & 0
 \end{pNiceMatrix} \begin{pNiceMatrix}
\tilde Q^{g\sdag}_{11}\\ \tilde Z^{g\sdag}_{11}
\end{pNiceMatrix}.
\eeq
We have denoted this basis as $\tilde Z^g_{11}$, since it is actually the analogue of $\tilde Z_{11}$ of proper eigenvectors. Indeed, the projected $P_+^\sdag\Phi^g_0$ will be given at step $m$ by $P^\sdag_+\Phi^g_0 = \tilde Z^g_{11}\tilde R^g_{11}$ for some matrix $\tilde R^g_{11}$. The recursion should run until $P^\sdag_+\Phi^g_0$, together with the projected proper eigenvectors $P^\sdag_+\Phi^r_0 = \tilde Z_{11}\tilde R_{11}$ (see Eq. \eqref{Z11}), form a complete basis of the $P_+$ subspace. To determine this, we must take care of the possibility that the $\tilde Z^g_{11}$ basis could end up including some strongly decaying \emph{proper} eigenvectors already present in $\tilde Z_{11}$ due to rounding errors. The best strategy to find the the rank of the combined basis $\begin{pNiceMatrix} \tilde Z_{11} & \tilde Z^g_{11}\end{pNiceMatrix}$ is once more to use a pivoted LQ decomposition to filter out redundant basis vectors,
\beq
\label{PZ}
\begin{pNiceMatrix} \tilde Z_{11} & \tilde Z^g_{11}\end{pNiceMatrix} = 
\begin{pNiceMatrix}[
	first-row,
	code-for-first-row = \scriptstyle]
r_1 & r_2\\
L_Z & 0
 \end{pNiceMatrix} \begin{pNiceMatrix}
P_Z^\sdag\\ \bar P_Z^\sdag
\end{pNiceMatrix}.
\eeq
As soon as $L_Z$ is a full-rank square matrix (i.e. $r_1 = r$), we have a complete basis $\begin{pNiceMatrix} \tilde Z_{11} & \tilde Z^g_{11}\end{pNiceMatrix}P_Z=L_Z$ of the $P_+$ subspace, spanned by proper plus generalized eigenvectors without redundancies, and we may stop the recursion. We can then express $P_+^\sdag\begin{pNiceMatrix} \Phi^r & \Phi^g_0\end{pNiceMatrix}$  in Eq. \eqref{Sigma} as
\beq
P_+^\sdag\begin{pNiceMatrix} \Phi^r & \Phi^g_0\end{pNiceMatrix} = \begin{pNiceMatrix} \tilde Z_{11} & \tilde Z^g_{11}\end{pNiceMatrix}
\overbrace{\begin{pNiceMatrix} \tilde R_{11} & 0 \\ 0 & \tilde R^g_{11}\end{pNiceMatrix}}^{R_{11}} =  L_Z P_Z^\sdag R_{11}
\eeq

Similarly, $h_-\Phi^g_0J^r_0 = P_+\tilde \Sigma^m_{11} P^\sdag_+\Phi^g_0 = P_+\tilde \Sigma^m_{11} \tilde Z^g_{11}\tilde R^g_{11}$, or equivalently
\beq
\begin{pNiceMatrix} h_-\Phi^r\Lambda^r & h_-\Phi^g_0J^r_0\end{pNiceMatrix} = 
\begin{pNiceMatrix} h_- Q_+\tilde Z_{21} & P_+ \tilde\Sigma^m_{11} \tilde Z_{11}^g\end{pNiceMatrix}
R_{11}
\eeq
Inserting the above results in Eq. \eqref{Sigma}, $R_{11}$ once more drops out, and we arrive at
\beq
\label{Sigmafinal}
\Sigma^r_{11} =  \begin{pNiceMatrix} h_- Q_+\tilde Z_{21} & P_+ \tilde\Sigma^m_{11} \tilde Z_{11}^g\end{pNiceMatrix}P_Z L_Z^{-1}P_+^\sdag
\eeq

This equation is the most general result of this work. To summarize the generalized eigenvector reconstruction procedure of this section, we start from Eq. \eqref{Tfinal}. The method has five basic steps: (1) build $\tilde g_0$, $\tilde t_\pm$ and $\tilde \Sigma_0$ of Eq. \eqref{tilde}, (2) iterate Eq. \eqref{recursive2} starting from $\tilde \Sigma^1_{11} = \tilde \Sigma_0$, (3) compute the kernel basis $\tilde Z^g_{11}$ of Eq. \eqref{sigmakernel} at each iteration, (4) combine it with the $\tilde Z_{11}$ of Eq. \eqref{Tfinal} to find $L_Z$ and $P_Z$ using Eq. \eqref{PZ}, and (5) as soon as we obtain a full rank (square) $L_Z$ matrix, stop the iteration, store the last $\tilde Z^g_{11}, \tilde\Sigma^m$, and insert all the ingredients in Eq. \eqref{Sigmafinal}.

\section{Benchmarks}

The combined deflate plus reconstruction procedure yielding Eq. \eqref{Sigmafinal} constitutes the equivalent to Eq. \eqref{T2} for $\Sigma^r_{11} = h_-T_r$, albeit computed fully trough manipulations of deflated matrices of size $r\times r$ or $2r\times 2r$ instead of the $2n\times 2n$ matrices of Eq. \eqref{T2}. Here, recall, $r$ is the dimension of orbitals on the surface of the unit cell, and $n$ is the total number of orbitals in the unit cell. 

An implementation of the proposed deflation+reconstruction Schur algorithm is available in the open-source package \verb|Quantica.jl|\cite{Quantica:Z21}, a MIT-licensed quantum simulation library written in the Julia language. It also implements the original Schur algorithm. In this section we use \verb|Quantica.jl| to compare the runtime performance of the two methods as a function of $n$ and $r$. To this end we build a quasi-1D graphene nanoribbon of width $W$ and lattice constant $a_0$. Its Hamiltonian is modeled using a nearest-neighbor tight-binding description. We assume the nanoribbon is `chiral', with edges along an axis between zig-zag and armchair orientations. Its translation symmetry is given by the Bravais vector $\bm{a} = m_1 \bm{a}_1 + m_2 \bm{a}_2$, where $m_{1,2}$ are integer chiral indices and $\bm{a}_{1,2}$ are graphene's Bravais vectors.

\begin{figure}
   \centering
   \includegraphics[width=\columnwidth]{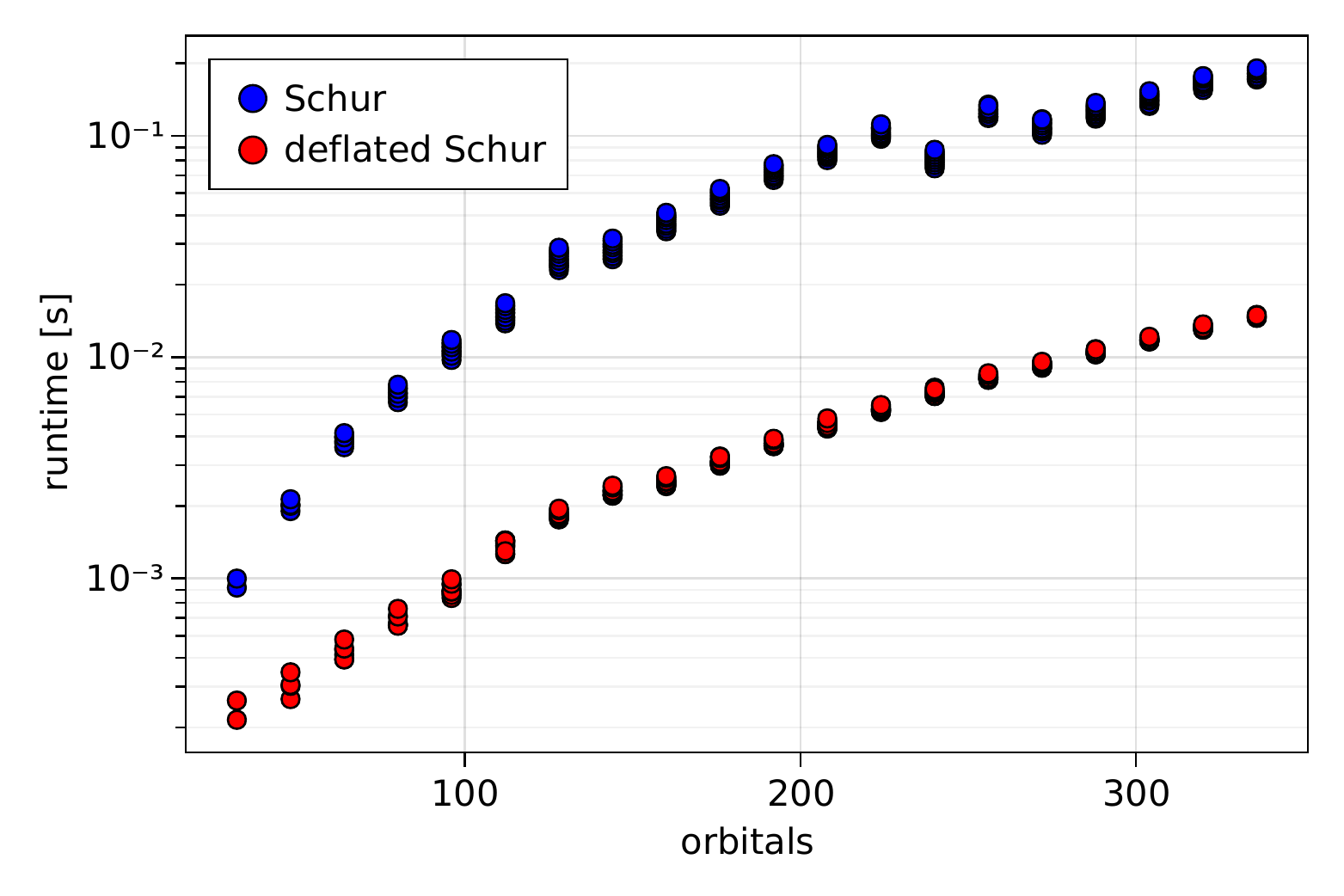}\\
    \includegraphics[width=\columnwidth]{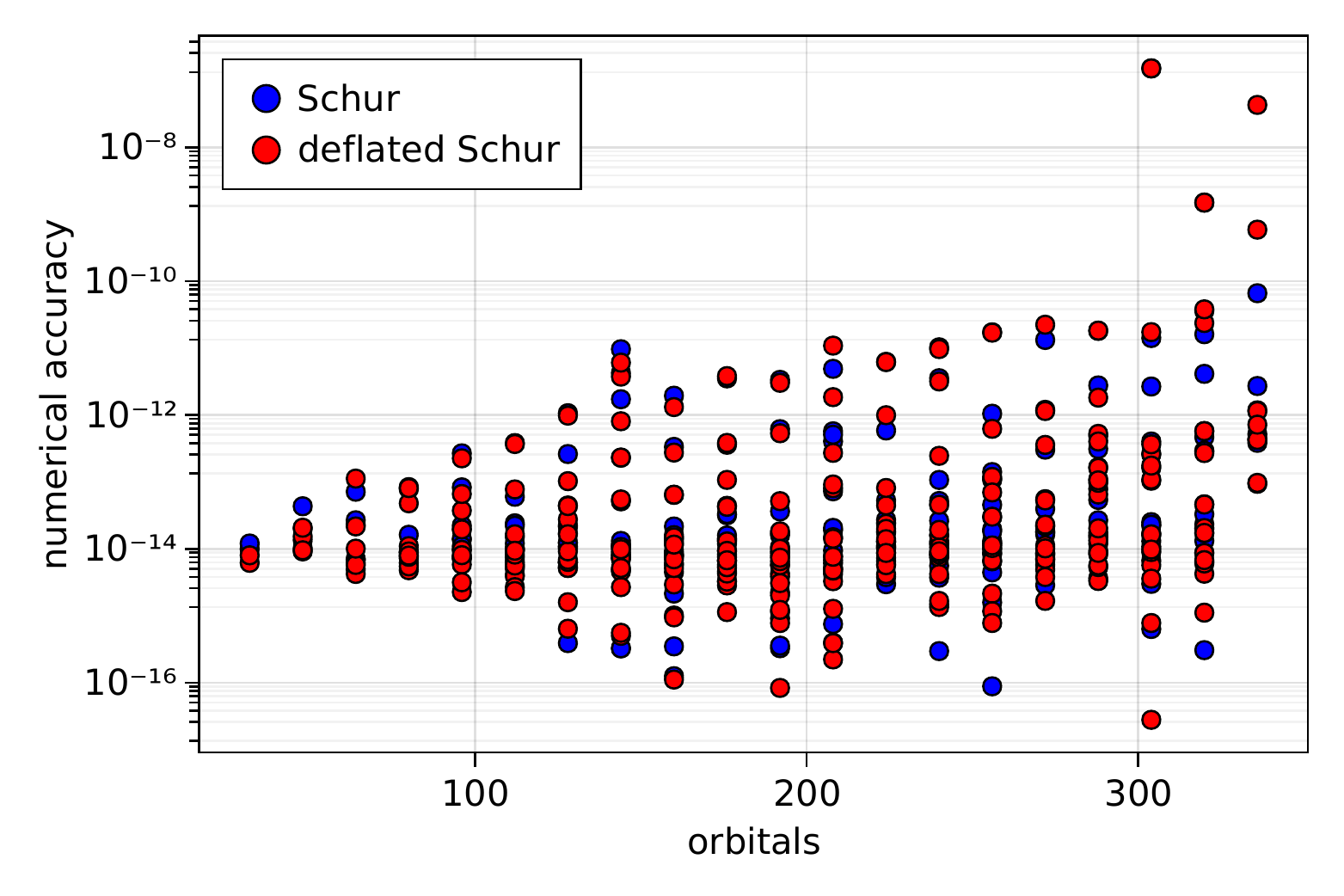}\\
   \caption{Runtime (top) and numerical accuracy (bottom) for one evaluation of the Green function $G^r_{11}(\omega)$ in different semi-infinite chiral graphene nanoribbons with an increasing number of orbitals per unit cell, using the standard Schur algorithm (blue) and the deflated version in this paper (red).}
   \label{fig:bench}
\end{figure}

\begin{figure*}
   \centering
   \includegraphics[width=\textwidth]{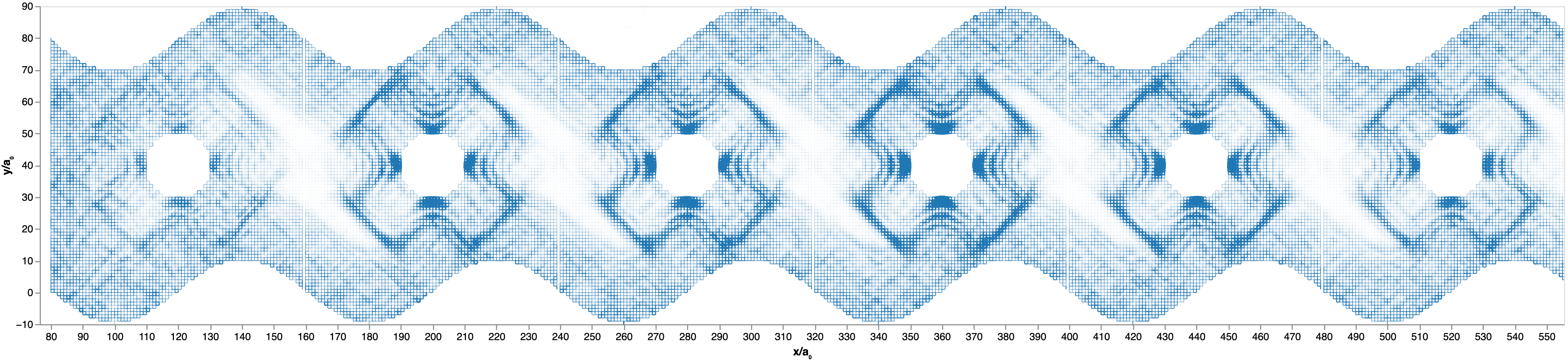}
   \caption{Example of a local density of states calculation in an elaborate, semi-infinite nanoribbon, computed on each unit cell $n$  using the deflation algorithm. Each unit cell is comprised of 6099 orbitals, which is reduced to 80 through deflation.}
   \label{fig:ldos}
\end{figure*}

Figure \ref{fig:bench} shows, in log-scale, some runtime results versus number of orbitals $n$ in the unit cell using the original and the deflated Schur methods. Each point corresponds to a single evaluation of $G^r_{11}(\omega)$ in nanboribbons with increasing $n$. All nanoribbons have a fixed width $W=7a_0$ in the $y$ direction and different $m_{1,2}$. The timings were measured on a desktop with an Intel Xeon X5650 processor, running Julia v1.7 and Quantica.jl v0.5.0, using double precision (64 bit) floating point arithmetic. We find that in this particular case the deflation step reduces runtime by a factor $\sim 10-20$, including deflation and reconstruction times. As shown also in Fig. \ref{fig:bench}, the numerical accuracy of both methods is essentially the same. We compute the accuracy by extracting the maximum entry in absolute value of the residual matrix
\beq
\rho = \Sigma^r_{11} - h_-\left(g_0^{-1} - \Sigma^r_{11}\right)^{-1}h_+,
\eeq
which should be identically zero for an exact solution $\Sigma^r_{11}$, see Eq. \eqref{G}.

Finally we showcase an example calculation where the non-deflated Schur becomes problematically slow, and which is solved efficiently using the deflation strategy. It corresponds to a square-lattice nanoribbon of width $W=80 a_0$ and sinusoidal edges of period $L=W$. Each period is decorated with a circular hole. Each site is connected to its four neighbors by a hopping $t=1$, onto which a Peierls phase is applied representing a magnetic flux just strong enough to drive the ribbon into the Quantum Hall regime. The local density of states $\rho_n = -\mathrm{Im}\,\mathrm{Tr}\,G_{nn}(\omega)$ is evaluated at $\omega = 0.2 t$. The result, computed with \verb|Quantica.jl| using the deflating algorithm is shown in Fig. \eqref{fig:ldos}. The unit cell contains 6099 orbitals, which is deflated to just 80. We find that the result is just as stable and accurate as with the undeflated Schur method, while being many times faster.

\section{Conclusions}

We have shown that it is possible to optimize the single-shot Schur algorithm for the computation of Green functions on a quasi-1D periodic lattice, without sacrificing its numerical stability properties. In essence our deflation method revolves around identifying the kernel of the intercell hopping matrices, and using it to filter out whole unwanted eigenvector subspaces of the transfer matrix $T^r$. To make the method work in general, it becomes crucial to pay attention to generalized eigenstates that are not obtained from the solutions to the quadratic eigenvalue equation for $T^r$. We develop a reconstruction post-processing step that restores the full spectral structure of the retarded Green function, including the generalized $\lambda=0$ eigenvectors, lost to deflation, that represent fully confined (non-evanescent) surface states.

\acknowledgements

We are grateful to Fernando Pe\~naranda for fruitful discussions and for critically reading this manuscript. This work was funded by the Spanish Ministry of Science, Innovation and Universities, Grant Nos. PCI2018-093026 and PGC2018-097018-B-I00 (AEI/FEDER).

\appendix

\section{Green functions between different unit cells}
\label{ap:GNM}

The Green function $G^r_{NM}$ from unit cell $M$ to unit cell $N$ in a semi-infinite lattice towards the right ($N,M>0$, with $N,M=1$ corresponding to the first unit cell) can be computed using only $G^r_{11} = (g_0^{-1}-\Sigma^r_{11})^{-1}$ and $\bar{G}^r_{11} = (g_0^{-1}-\bar\Sigma^r_{11})^{-1}$, where $\Sigma^r_{11} = h_-G^r_{11}h_+$, $\bar\Sigma^r_{11} = h_+\bar G^r_{11}h_-$ and $\bar G^r_{11}$ is the Green function of a semi-infinite lattice with $h_-$ and $h_+$ interchanged. In this section we give general expressions.

The special case with $M=1$ (i.e. propagation starting on the first unit cell) admits a simpler solution. By using the Dyson equation we can write $G^r_{N,1} = G^r_{N-1,1}h_+G^r_{11}$. Solving this recursively we get
\beq
\label{GN1}
G^r_{N1} = (G^r_{11}h_+)^{N-1}G^r_{11} = (T^r)^{N-1}G^r_{11}
\eeq

The physical interpretation of this is clear. The $G^r_{11}$ matrix propagates within the leftmost unit cell in the presence of the whole semi-infinite lattice, while the transfer matrix operator $T^r$ propagates one cell to the right. The above expression can be re-expressed using the intra-unit-cell retarded propagator of the \emph{infinite} (not semi-infinite) lattice,
\beq
G^{\infty}_{00} = (g_0^{-1}-\Sigma^r_{11}-\bar\Sigma^r_{11})^{-1}
\eeq
We have
\beq
G^r_{N1} = \left[(T^r)^{N-1} - (T^r)^{N}\bar T^r\right]G^{\infty}_{00}
\eeq
Here $\bar T^r = \bar G^r_{11}h_-$. The equivalence of both expressions for $G^r_{N1}$ is easy to verify by multiplying with $(G^{\infty}_{00})^{-1} = (\bar G^r_{11})^{-1} - \bar\Sigma^r_{11}$ on the right. The second relation, however, allows for a valuable interpretation. The effect of cutting the infinite lattice at cell $N=0$, turning it into two semi-infinite halves, is to give the propagator $G^r_{N1}$ two contributions, one traveling from cell 1 to N $(T^r)^{N-1}G^{\infty}_{00}$, and another traveling leftwards to the removed cell at $M=0$ that is then reflected rightwards back to $N$, i.e. $(T^r)^{N}\bar T^rG^{\infty}_{00}$. This is reminiscent also of the method of images. It allows us to extrapolate the general expression for $G^r_{NM}$,
\beq
G^r_{NM} = \left[(T^r)^{N-M} - (T^r)^{N}(\bar T^r)^M\right]G^{\infty}_{00}.
\eeq

\section{Alternative deflation algorithm}
\label{ap:quadeig}

In Sec. \ref{sec:deflation} we have developed a deflation algorithm based on the first companion linearization $C_1$ of the quadratic eigenvalue equation Eq. \ref{lin}. The specific linearization is defined as
\beq
\label{C1}
C_1:\space \Bigg[\overbrace{\left(\begin{array}{cc}
0 & 1\\ -h_+ & g_0^{-1}
\end{array}\right)}^{A}
-
\lambda\overbrace{\left(\begin{array}{cc}
1 &0 \\ 0 &h_-
\end{array}\right)}^{B}\Bigg]
\overbrace{\left(\begin{array}{c}
\phi \\ \lambda\phi
\end{array}\right)}^{\psi} = 0
\eeq

An important step in the algorithm required that the kernel of $h_+$ and $h_-^\sdag$ be the same, see Eqs. \eqref{h+} and \eqref{h-}. While this is the case in most problems of interest, and in particular in Hermitian ones where $h_+=h_-^\sdag$, one can envision models for problems in which $h_+$ and $h_-^\sdag$ are completely independent matrices, with different kernels, even different rank. In such cases a more elaborate deflation algorithm, as the one sketched in this section, becomes necessary. 

The alternative deflation scheme build on the \verb|quadeig| algorithm \cite{Hammarling:ATMS13}, adapted to work in tandem with the Schur method. The \verb|quadeig| algorithm starts with a different linearization, so-called second companion $C_2$, defined as
\beq
\label{C2}
C_2:\space \Bigg[\overbrace{\left(\begin{array}{cc}
g_0^{-1} & -1\\ -h_+ & 0
\end{array}\right)}^{A}
-
\lambda\overbrace{\left(\begin{array}{cc}
h_- &0 \\ 0 &-1
\end{array}\right)}^{B}\Bigg]
\overbrace{\left(\begin{array}{c}
\phi \\ \lambda^{-1}h_+\phi
\end{array}\right)}^{\psi} = 0
\eeq

By a sequence of transformations, explained in full detail in Ref. \cite{Hammarling:ATMS13}, the \verb|quadeig| algorithm transforms $A$ and $B$ into upper-block-triangular form, with the lower blocks corresponding to $\lambda=0$ and $\lambda=\infty$ solutions that can be thus dropped, leaving only the upper left block as a deflated version of $A, B$, to compute retarded and advanced finite eigenvalues, much like in our Eq. \eqref{defpencil}. No assumption is made about the structure of $h_+$ or $h_-$. A crucial difference with our deflation scheme is that the lower block of the linearized $\psi$ eigenvectors in $C_2$ above is actually $\chi=\lambda^{-1} h_+\phi$, where $\phi$ is a proper $T^r$ eigenvector. A key mathematical advantage of the Schur algorithm is that it is able to use the invariant subspace basis $Z_{21}$ of this lower block to construct $T^r$ without needing to compute or invert the eigenstates themselves. This property is spoilt by the $h_+$ in $\chi$ for $C_2$. The Schur algorithm with a deflated $C_2$ could be adapted to produce $h_+G^r_{11}h_+$, which unlike $T^r=G^r_{11}h_+$ is insufficient to reconstruct $G^r_{11}$.

This analysis, however, suggests a solution. If instead of the second we use the fourth companion linearization $C_4$
\beq
\label{C4}
C_4:\space \Bigg[\overbrace{\left(\begin{array}{cc}
-h_+ &0 \\ 0 &1
\end{array}\right)}^{A}
-
\lambda\overbrace{\left(\begin{array}{cc}
-g_0^{-1} & 1\\ h_+ & 0
\end{array}\right)}^{B}\Bigg]
\overbrace{\left(\begin{array}{c}
\phi \\ \lambda h_-\phi
\end{array}\right)}^{\psi} = 0
\eeq
we can see that the form of $\chi = \lambda h_-\phi$ now allows us to build $\Sigma^r_{11} = h_- G^r_{11}h_+ = Z_{21}Z_{11}^{-1}$ following the same derivation of Eq. \eqref{T2}. With $\Sigma^r_{11}$ it is then straightforward to obtain $G^r_{11}$ through Eq. \eqref{G}. The problem then reduces to adapt the \verb|quadeig| algorithm to the $C_4$ linearization. This is easy however, since $C_4$ is related to $C_2$ by $A\leftrightarrow B$ and $h_+\leftrightarrow h_-$. One therefore just needs to build $C_2$ with $h_\pm$ replaced by $h_\mp$, deflate with the \verb|quadeig| method, and finally interchange the deflated $A$, $B$ before proceeding with the Schur method and the generalized eigenstate reconstruction. The final Eq. \ref{Sigmafinal} would be identical, except that $h_-Q_+\tilde Z_{21}$ should read simply $Q_+\tilde Z_{21}$, as $h_-$ is already incorporated into $Z_{21}$. 

Although slightly more involved to implement, this modified \verb|quadeig| algorithm is similar in complexity to the deflation procedure in Sec. \ref{sec:deflation}. Both are much more performant than older deflation strategies, such as e.g. Van Dooren algorithm \cite{Van-Dooren:LAAIA79}, that do not exploit the structure of the linearization matrices, which is ultimately what allows us to eliminate both $\lambda=0$ and $\lambda=\infty$ subspaces in the same pass. Both, moreover, can make use of scaling and balancing prior to deflation to maximize numerical accuracy \cite{Fan:SJMAA04,Hammarling:ATMS13,Drmac:19}, although such strategies are not explicitly discussed here.
 
\bibliography{/Users/pablo/Seafile/Pablo/Bibliography/BibdeskPablo/biblio}
\end{document}